\documentclass[article,superscriptaddress,twocolumn]{revtex4}
\usepackage[ansinew]{inputenc}
\usepackage{graphicx}
\usepackage{amsmath}
\usepackage{amsthm}
\usepackage{bm}
\usepackage{layout}
\usepackage{float}
\usepackage{txfonts}
\usepackage{amsfonts}
\usepackage{amssymb}%
\usepackage[ansinew]{inputenc}
\usepackage{graphicx}
\usepackage{amsmath}
\usepackage{amsthm}
\usepackage{bm}
\usepackage{layout}
\usepackage{float}
\usepackage{amsfonts}
\usepackage{amssymb}

\def\bp{\mathbf{p}}
\def\bb{\mathbf{b}}

\def\bE{\mathbf{E}}

\def\br{\mathbf{r}}

\def\bx{\mathbf{x}}
\def\by{\mathbf{y}}
\def\bz{\mathbf{z}}
\def\be{\mathbf{e}}
\def\bT{\mathbf{T}}
\def\bm{\mathbf{m}}

\def\cS{\mathcal{S}}
\def\ket#1{|#1\rangle}
\def\bra#1{\langle#1|}

\newtheorem{Axiom}{{\bf Axiom}}
\newtheorem{Lemma}{{\bf Lemma}}
\newtheorem{lemma}{{\bf Lemma}}
\newtheorem*{Axiom'}{{\bf Axiom 3'}}

\usepackage{bm}
\usepackage{graphicx}

\begin{document}

\title{Quantum Theory and Beyond: Is Entanglement Special?}

\author{Borivoje Daki\'c}
\affiliation{Faculty of Physics, University of Vienna,
Boltzmanngasse 5, A-1090 Vienna, Austria}
\author{{\v C}aslav Brukner}
\affiliation{Faculty of Physics, University of Vienna,
Boltzmanngasse 5, A-1090 Vienna, Austria}
\affiliation{Institute of Quantum Optics and Quantum Information,
Austrian Academy of Sciences, Boltzmanngasse 3, A-1090 Vienna,
Austria}
\begin{abstract}

Quantum theory makes the most accurate empirical predictions and yet
it lacks simple, comprehensible physical principles from
which the theory can be uniquely derived. A broad class of
probabilistic theories exist which all share some features with
quantum theory, such as probabilistic predictions for individual
outcomes (indeterminism), the impossibility of information transfer
faster than speed of light (no-signaling) or the impossibility of
copying of unknown states (no-cloning). A vast majority of attempts
to find physical principles behind quantum theory either fall short of
deriving the theory uniquely from the principles or are based on abstract mathematical
assumptions that require themselves a more conclusive physical motivation.
Here, we show that classical probability theory and quantum
theory can be reconstructed from three reasonable axioms: (1)
(Information capacity) All systems with information carrying capacity
of one bit are equivalent. (2) (Locality) The state of a composite
system is completely determined by measurements on its subsystems.
(3) (Reversibility) Between any two pure states there exists a
reversible transformation. If one requires the transformation from
the last axiom to be continuous, one separates quantum theory from
the classical probabilistic one. A remarkable result following from
our reconstruction is that {\it no probability theory other than
quantum theory can exhibit entanglement without contradicting one or
more axioms.}

\end{abstract}

\maketitle

\section{Introduction}

The historical development of scientific progress teaches us that
every theory that was established and broadly accepted at a certain
time was later inevitably replaced by a deeper and more fundamental
theory of which the old one remains a special case. One celebrated
example is Newtonian (classical) mechanics which was superseded by
quantum mechanics at the beginning of the last century. It is
natural to ask whether in a similar manner there could be logically
consistent theories that are more generic than quantum theory
itself. It could then turn out that quantum mechanics is an
effective description of such a theory, only valid within our
current restricted domain of experience.

At present, quantum theory has been tested against very specific
alternative theories that, both mathematically and in their
concepts, are distinctly different. Instances of such alternative
theories are non-contextual hidden-variable theories~\cite{KS},
local hidden-variable theories~\cite{Bell}, crypto-nonlocal
hidden-variable theories~\cite{Leggett,Groeblacher}, or some
nonlinear variants of the Schrödinger
equation~\cite{Birula,Shimony,Shull,Gaehler}. Currently, many groups are working
on improving experimental conditions to be able to test alternative
theories based on various collapse
models~\cite{GWR,Karoli,Diosi,Penrose,Pearle}. The common trait of all these proposals
is to suppresses one or the other counter-intuitive feature of quantum mechanics and thus keep
some of the basic notions of a classical world view intact. Specifically, hidden-variable models would
allow to preassign definite values to outcomes of all measurements,
collapse models are mechanisms for restraining superpositions
between macroscopically distinct states and nonlinear extensions of
the Schrödinger equation may admit more localized solutions for
wave-packet dynamics, thereby resembling localized classical particles.

In the last years the new field of quantum information has
initialized interest in generalized probabilistic theories which
share certain features -- such as the no-cloning and the
no-broadcasting theorems~\cite{Barnum06,Barnum07} or the trade-off
between state disturbance and measurement~\cite{Barrett} --
generally thought of as specifically quantum, yet being shown to be
present in all except classical theory. These generalized probabilistic
theories can allow for stronger than quantum correlations in the sense
that they can violate Bell's inequalities stronger than the quantum
Cirel’son bound (as it is the case for the celebrated ``non-local
boxes'' of Popescu and Rohrlich~\cite{PRBox}), though they all
respect the ``non-signaling'' constraint according to which
correlations cannot be used to send information faster than the
speed of light.

Since the majority of the features that have been highlighted as
``typically quantum'' are actually quite generic for all
non-classical probabilistic theories, one could conclude that
additional principles must be adopted to single out quantum theory
uniquely. Alternatively, these probabilistic theories indeed can be constructed
in a logically consistent way, and might even be realized in nature
in a domain that is still beyond our observations. The vast majority
of attempts to find physical principles behind quantum theory either fail
to single out the theory uniquely or are based on highly abstract
mathematical assumptions without an immediate physical meaning (e.g.~\cite{Mackey}).

On the way to reconstructions of quantum theory from foundational
physical principles rather than purely mathematical axioms, one
finds interesting examples coming from an instrumentalist
approach~\cite{Hardy,Dariano,Goyal2}, where the focus is primarily
on primitive laboratory operations such as preparations,
transformations and measurements. While these reconstructions are
based on a short set of simple axioms, they still partially use
mathematical language in their formulation.

Evidentally,  added value of reconstructions for better understanding
quantum theory originates from its power of explanation where the
structure of the theory comes from. Candidates for foundational
principles were proposed giving a basis for an understanding of
quantum theory as a general theory of information supplemented by
several information-theoretic
constraints~\cite{Rovelli,Zeilinger,BruknerZeilinger,CBH,Grinbaum1}.
In a wider context these approaches belong to attempts to find an
explanation for quantum theory by putting primacy on the concept of
information or on the concept of probability which again can be seen
as a way of quantifying
information~\cite{Wootters,Fivel,Summhammer,Bohr,Caticha,Fuchs,Grangier,Luo,Spekkens,Goyal1}.
Other principles were proposed for separation of quantum
correlations from general non-signaling correlations, such as that
communication complexity is not trivial~\cite{Vandam,Brassard}, that
communication of $m$ classical bits causes information gain of at
most $m$ bits (``information causality'')~\cite{Pawlowski}, or that
any theory should recover classical physics in the macroscopic
limit~\cite{Navascues}.

In his seminal paper, Hardy~\cite{Hardy} derives quantum theory from
``five reasonable axioms'' within the instrumentalist framework. He
sets up a link between two natural numbers, $d$ and $N$,
characteristics of any theory. $d$ is the number of degrees of
freedom of the system and is defined as the minimum number of real
parameters needed to determine the state completely. The dimension
$N$ is defined as the maximum number of states that can be reliably
distinguished from one another in a single shot experiment. A
closely related notion is the information carrying capacity of the
system, which is the maximal number of bits encoded in the system,
and is equal to $\log N$ bits for a system of dimension $N$.

Examples of theories with an explicit functional dependence $d(N)$
are classical probability theory with the linear dependence $d=N-1$,
and quantum theory with the quadratic dependence for which it is
necessary to use $d=N^2-1$ real parameters to completely
characterize the quantum state~\footnote{Hardy considers
unnormalized states and for that reason takes $K=d+1$ (in his
notation) as the number of degrees of freedom.}. Higher-order
theories with more general dependencies $d(N)$ might exist as
illustrated in Figure 1. Hardy's reconstruction resorts to a
``simplicity axiom'' that discards a large class of higher-order
theories by requiring that for each given $N$, $d(N)$ takes the
minimum value consistent with the other axioms. However, without
making such an {\it ad hoc} assumption the higher-order theories
might be possible to be constructed in agreement with the rest of
the axioms. In fact, an explicit quartic theory for which $d=
N^4-1$~\cite{Zyczkowski}, and theories for generalized bit ($N=2$)
for which $d=2^r-1$ and $r\in N$~\cite{Paterek}, were recently
developed, though all of them are restricted to the description of individual systems only. 

\begin{figure}\centering
\includegraphics[width=9cm]{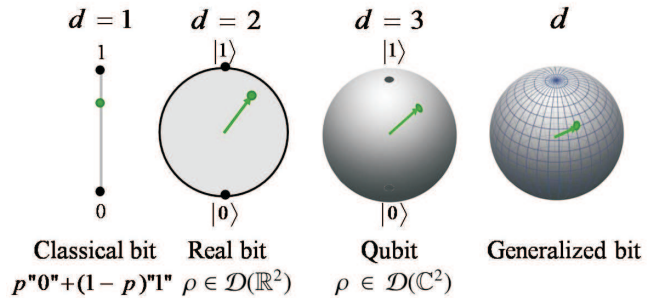}
\caption{State spaces of a two-dimensional system in the generalized
probabilistic theories analyzed here. $d$ is the minimal number of
real parameters necessary to determine the (generally mixed) state completely. From
left to right: A classical bit with one parameter (the weight $p$ in
the mixture of two bit values), a real bit with two real parameters
(state $\rho \in \mathcal{D}(\mathbb{R}^2)$ is represented by $2 \times 2$ real density matrix), a qubit (quantum bit) with three real
parameters (state $\rho \in \mathcal{D}(\mathbb{C}^2)$ is represented by $2 \times 2$ complex density matrix) and a
generalized bit for which $d$ real parameters are needed to specify
the state. Note that, when one moves continuously from one pure
state (represented by a point on the surface of a sphere) to
another, only in the classical probabilistic theory one must go
trough the set of mixed states. Can probability theories that are
more generic than quantum theory be extended in a logically
consistent way to higher-dimensional and composite systems? Can
entanglement exist in these theories? Where should we look in nature
for potential empirical evidences of the theories?}
\end{figure}

It is clear from the previous discussion that the question on basis of which
physical principles quantum theory can be separated from the
multitude of possible generalized probability theories is still
open. A particulary interesting unsolved problem is whether the
higher-order theories of Refs.~\cite{Hardy,Zyczkowski,Paterek} can
be extended to describe non-trivial, i.e. {\it entangled}, states of
composite systems. Any progress in theoretical understanding of
these issues would be very desirable, in particular because experimental
research efforts in this direction have been very sporadic. Although
the majority of experiments indirectly verify also the number of the
degrees of freedom of quantum systems~\footnote{As noted by
Zyczkowski~\cite{Zyczkowski} it is thinkable that within the time
scales of standard experimental conditions ``hyper-decoherence'' may
occur which cause a system described in the framework of the
higher-order theory to specific properties and behavior according to
predictions of standard (complex) quantum theory.}, there are only
few dedicated attempts at such a direct experimental verification.
Quaternionic quantum mechanics (for which $d=2N^2-N-1$) was tested
in a suboptimal setting~\cite{Peres2} in a single neutron experiment
in 1984~\cite{Peres1,Kaiser}, and more recently, the generalized
measure theory of Sorkin~\cite{Sorkin} in which higher order
interferences are predicted was tested in a three-slit experiment
with photons~\cite{Weihs}. Both experiments put an upper bound on
the extent of the observational effects the two alternative theories
may produce.

\section{Basic Ideas and the Axioms}

Here we reconstruct quantum theory from three reasonable axioms.
Following the general structure of any reconstruction we first give
a set of physical principles, then formulate their mathematical
representation, and finally rigorously derive the formalism of the
theory. We will only consider the case where the number of
distinguishable states is finite. The three axioms which separate
classical probability theory and quantum theory from all other
probabilistic theories are:

\begin{Axiom}\label{Axiom: capacity}
(Information capacity) An elementary system has the information
carrying capacity of at most one bit. All systems of the same
information carrying capacity are equivalent.
\end{Axiom}
\begin{Axiom}\label{Axiom:local} (Locality)
The state of a composite system is completely determined by local
measurements on its subsystems and their correlations.
\end{Axiom}
\begin{Axiom}\label{Axiom: equivalence}
(Reversibility) Between any two pure states there exists a reversible
transformation.
\end{Axiom}

A few comments on these axioms are appropriate here. The most
elementary system in the theory is a two-dimensional system. All
higher-dimensional systems will be built out of two-dimensional
ones. Recall that the dimension is defined as the maximal number of
states that can be reliably distinguished from one another in a
single shot experiment. Under the phrase  ``an elementary system has
an information capacity of at most one bit'' we precisely assume
that for any state (pure or mixed) of a two-dimensional system there
is a measurement such that the state is a mixture of {\em two}
states which are distinguished reliable in the measurement. An
alternative formulation could be that {\it any state of a two
dimensional system can be prepared by mixing at most two basis (i.e.
perfectly distinguishable in a measurement) states} (see Figure 2).
Roughly speaking, axiom 1 assumes that a state of an elementary
system can always be represented as a mixture of {\it two classical
bits}. This part of the axiom is inspired by Zeilinger's proposal for a foundation principle for quantum theory~\cite{Zeilinger}.

\begin{figure}\centering
\includegraphics[width=6.5cm]{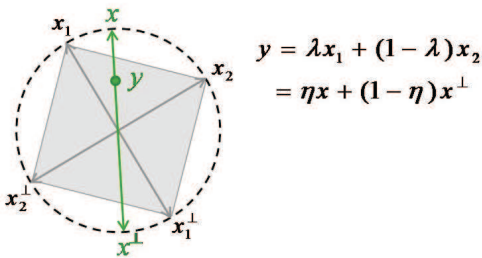}
\caption{Illustration of the assumption stated in axiom 1. Consider
a toy-world of a two-dimensional system in which the set of pure
states consists of only $\bx_1$ and $\bx_2$ and their orthogonal
states $\bx_1^{\perp}$ and $\bx_2^{\perp}$ respectively, and where
only two measurements exist, which distinguish
$\{\bx_1,\bx_1^{\perp}\}$ and $\{\bx_2,\bx_2^{\perp}\}$. The convex
set (represented by the grey area within the circle) whose vertices
are the four states contains all physical (pure or mixed) states in
the toy-world. Now, choose a point in the set, say $\by=\lambda
\bx_1 + (1-\lambda) \bx_2$. Axiom 1 states that any physical state
can be represented as a mixture of two orthogonal states (i.e.
states perfectly distinguishable in a single shot experiment), e.g.
$\by=\eta \bx + (1-\eta) \bx^{\perp}$. This is not fulfilled in the
toy world, but is satisfied in a theory in which the entire circle
represents the pure states and where measurements can distinguish
all pairs of orthogonal states.}\label{Decomposition}
\end{figure}

The second statement in axiom 1 is motivated by the intuition that
at the fundamental level there should be no difference between
systems of the same information carrying capacity. All elementary
systems -- be they part of higher dimensional systems or not --
should have equivalent state spaces and equivalent sets of
transformations and measurements. This seems to be a natural
assumption if one makes no prior restrictions to the theory and
preserves the full symmetry between all possible elementary systems.
This is why we have decided to put the statement as a part of axiom
1, rather than as a separate axiom. The particular formulation used
here is from Grinbaum~\cite{Grinbaum2} who suggested to rephrase the
``subspace axiom'' of Hardy's reconstruction using physical language
rather than mathematical. The subspace axiom states that a system
whose state is constrained to belong to an $M$ dimensional subspace
(i.e. have support on only $M$ of a set of $N$ possible
distinguishable states) behaves like a system of dimension $M$.

In logical terms axiom 1 means the following. We can think of two
basis states as two binary propositions about an individual system,
such as (1) ``The outcome of measurement $A$ is +1'' and (2) ``The
outcome of measurement $A$ is -1''. An alternative choice for the
pair of propositions can be propositions about joint properties of
two systems, such as (1') ``The outcomes of measurement $A$ on the
first system and of $B$ on the second system are correlated'' (i.e.
either both +1 or both -1) and (2') ``The outcomes of measurement
$A$ on the first system and of $B$ on the second system are
anticorrelated''. The two choices for the pair of propositions
correspond to two choices of basis states which each can be used to
span the full state space of an abstract elementary system (also
called ``generalized bit''). As we will see later, taking the latter
choice, it will follow from axiom 1 alone that the state space must
contain entangled states.

Axiom 2 assumes that a specification of the probabilities for a
complete set of local measurements for each of the subsystems plus
the joint probabilities for correlations between these measurements
is sufficient to determine completely the global state. Note that
this property does hold in both quantum theory and classical
probability theory, but not in quantum theory formulated on the
basis of real or quaternionic amplitudes instead of complex. A
closely related formulation of the axiom was given by
Barrett~\cite{Barrett}.

Finally, axiom 3 requires that transformations are reversible. This
is assumed alone for the purposes that the set of transformations
builds a group structure. It is natural to assume that a composition
of two physical transformations is again a physical transformation.
It should be noted that this axiom could be used to exclude the
theories in which ``non-local boxes'' occur, because there the
dynamical group is trivial, in the sense that it is generated solely
by local operations and permutations of systems with no entangling
reversible transformations (that is, non-local boxes cannot be
prepared from product states)~\cite{Gross}.

If one requires the reversible transformation from our axiom 3 to be
continuous:
\begin{Axiom'} 
(Continuity) Between any two pure states there exists a continuous reversible
transformation,
\end{Axiom'}

\hspace{-0.475cm} which separates quantum theory from classical
probability theory. The same axiom is also present in Hardy's
reconstruction. By a continuous transformation is here meant that
every transformation can be made up from a sequence of
transformations only infinitesimally different from the identity.

A remarkable result following from our reconstruction is that {\it
quantum theory is the only probabilistic theory in which one can
construct entangled states and fulfill the three axioms}. In
particular, in the higher-order theories of
Refs.~\cite{Hardy,Zyczkowski,Paterek} composite systems can only
enjoy trivial separable states. On the other hand, we will see that
axiom 1 alone requires entangled states to exist in all
non-classical theories. This will allow us to discard the
higher-order theories in our reconstruction scheme without invoking
the simplicity argument.

As a by product of our reconstruction we will be able to answer why 
in nature only ``odd'' correlations (i.e. $(1,1,-1)$, $(1,-1,1)$, $(-1,1,1)$
and $(-1,-1,-1))$  are observed when two maximally entangled qubits (spin-1/2 particles) are both measured along direction $x$, $y$ and $z$, respectively. The most familiar example is of the singlet state $|\psi^-\rangle=\frac{1}{2}(|0\rangle_1|1\rangle_2 - |0\rangle_1|1\rangle_2)$ with anticorrelated results for arbitrarily but the same choice of measurement directions for two qubits. We will show that the ``mirror quantum mechanics'' in which only ``even'' correlations appear cannot be extended  consistently to composite systems of three bits.

Our reconstruction will be given in the framework of typical
experimental situation an observer faces in the laboratory. While
this instrumentalist approach is a useful paradigm to work with, it
might not be necessary. One could think about axioms 1 and 3 as
referring to objective features of elementary constituents of the
world which need not necessarily be related to laboratory actions.
In contrast, axiom 2 seems to acquire a meaning only within the
instrumentalist approach as it involves the word ``measurement''.
Even here one could follow a suggestion of Grinbaum~\cite{Grinbaum2}
and rephrase the axiom to the assumption of ``multiplicability of
the information carrying capacity of subsystems.''

Concluding this section, we note that the conceptual groundwork for
the ideas presented here has been prepared most notably by
Weizsäcker~\cite{Weizsaecker}, Wheeler~\cite{Wheeler} and
Zeilinger~\cite{Zeilinger} who proposed that the notion of the
elementary yes-no alternative,  or the ``Ur'', should play a pivotal
role when reconstructing quantum physics.

\section{Basic notions}

Following Hardy~\cite{Hardy} we distinguish three types of devices
in a typical laboratory.  The preparation device prepares systems in
some state. It has a set of switches on it for varying the state
produced. After state preparation the system passes through a
transformation device. It also has a set of switches on it for
varying the transformation applied on the state. Finally, the system
is measured in a measurement apparatus. It again has switches on it
with which help an experimenter can choose different measurement
settings. This device outputs classical data, e.g. a click in a
detector or a spot on a observation screen.

We define the \emph{state} of a system as that mathematical object
from which one can determine the probability for any conceivable
measurement. Physical theories can have enough structure that it is
not necessary to give an exhaustive list of all probabilities for
all possible measurements, but only a list of probabilities for some
minimal subset of them. We refer to this subset as \emph{fiducial}
set. Therefore, the state is specified by a list of $d$ (where $d$ depends on dimension $N$)
probabilities for a set of fiducial measurements:
$\bp=(p_1,\dots,p_{d})$. The state is \emph{pure} if it is
not a (convex) mixture of other states. The state is mixed if it is
not pure. For example, the mixed state $\bp$ generated by preparing
state $\bp_1$ with probability $\lambda$ and $\bp_2$ with
probability $1-\lambda$, is $\bp=\lambda\bp_1+(1-\lambda)\bp_2$.

When we refer to an $N$-dimensional system, we assume that there are
$N$ states each of which identifies a different outcome of some
measurement setting, in the sense that they return probability one
for the outcome. We call this set a set of \emph{basis} or
\emph{orthogonal states}. Basis states can be chosen to be pure. To
see this assume that some mixed state identifies one outcome. We can
decompose the state into a mixture of pure states, each of which has
to return probability one, and thus we can use one of them to be a
basis state. We will show later that each pure state corresponds to
a unique measurement outcome.

If the system in state $\bp$ is incident on a transformation device,
its state will be transformed to some new state $U(\bp)$. The
transformation $U$ is a linear function of the state $\bp$ as it
needs to preserve the linear structure of mixtures. For example,
consider the mixed state $\bp$ which is generated by preparing state
$\bp_1$ with probability $\lambda$ and $\bp_2$ with probability
$1-\lambda$. Then, in each single run, either $\bp_1$ or $\bp_2$ is
transformed and thus one has:
\begin{equation}\label{linearity}
U(\lambda\bp_1+(1-\lambda)\bp_2)=\lambda
U(\bp_1)+(1-\lambda)U(\bp_2).
\end{equation}

It is natural to assume that a composition of two or more
transformations is again from a set of (reversible) transformations.
This set forms some abstract group.  Axiom 3 states that the
transformations are reversible, i.e. for every $U$ there is an
inverse group element $U^{-1}$. Here we assume that every
transformation has its matrix representation $U$ and that there is
an orthogonal representation of the group: there exists an
invertible matrix $S$ such that $O=SUS^{-1}$ is an orthogonal
matrix, i.e. $O^{\mathrm{T}}O=\openone$, for every $U$ (We use the same
notation both for the group element and for its matrix
representation). This does not put severe restrictions to the group
of transformations, as it is known that all compact groups have such
a representation (the Schur-Auerbach lemma)~\cite{Boerner}. Since
the transformation keeps the probabilities in the range $[0,1]$, it
has to be a compact group~\cite{Hardy}. All finite groups and all
continuous Lie groups are therefore included in our consideration.

Given a measurement setting, the outcome probability
$P_{\mathrm{meas}}$ can be computed by some function $f$ of the
state $\bp$,
\begin{equation}
P_{\mathrm{meas}}=f(\bp).
\end{equation}
Like a transformation, the measurement cannot change the mixing
coefficients in a mixture, and therefore the measured probability is
a linear function of the state $\bp$:
\begin{equation}
\label{linearity_measur}
f(\lambda\bp_1+(1-\lambda)\bp_2)=\lambda
f(\bp_1)+(1-\lambda)f(\bp_2).
\end{equation}

\section{Elementary system: system of information capacity of 1 bit}

A two-dimensional system has two distinguishable outcomes which can be
identified by a pair of basis states $\{\bp,\bp^{\perp}\}$. The state is specified by $d$ probabilities $\bp=(p_1,...,p_d)$ for $d$ fiducial measurements, where $p_i$ is probability for a particular outcome of the $i$-th fiducial measurement (the dependent probabilities $1-p_i$ for the opposite outcomes are omitted in the state description). Instead
of using the probability vector $\bp$ we will specify the state by its \emph{Bloch
representation} $\bx$ defined as a vector with $d$
components:
\begin{equation}\label{Bloch}
x_i=2p_i-1.
\end{equation}
The mapping between the two different representations is an invertible
linear map and therefore preserves the structure of the mixture
$\lambda\bp_1+(1-\lambda)\bp_2\mapsto\lambda\bx_1+(1-\lambda)\bx_2$.

It is convenient to define a \emph{totaly mixed state}
$\bE=\frac{1}{\mathcal{N}}\sum_{x\in\cS_{\mathrm{pure}}}\bx$, where
$\cS_{\mathrm{pure}}$ denotes the set of pure states and
$\mathcal{N}$ is the normalization constant. In the case of a
continuous set of pure states the summation has to be replaced by a
proper integral. It is easy to verify that $\bE$ is a totally
invariant state. This implies that every measurement and in
particular the fiducial ones will return the same probability for
all outcomes. In the case of a two-dimensional system this
probability is $1/2$. Therefore, the Bloch vector of the totally
mixed state is the zero-vector $\bE=\vec{0}$.

The transformation $U$ does not change the totaly mixed state, hence
$U(\vec{0})=\vec{0}$. The last condition together with the linearity
condition (\ref{linearity}) implies that any transformation is
represented by some $d\times d$ real invertible matrix $U$. The same
reasoning holds for measurements. Therefore, the measured
probability is given by the formula:
\begin{equation}\label{prob rule}
P_{\mathrm{meas}}=\frac{1}{2}(1+\br^{\mathrm{T}}\bx).
\end{equation}
The vector $\br$ represents the outcome for the given measurement
setting. For example, the vector $(1,0,0,\dots)$ represents one of
the outcomes for the first fiducial measurement.

According to axiom 1 any state is a classical mixture of some pair
of orthogonal states. For example, the totally mixed state is an
equally weighted mixture of some orthogonal states
$\vec{0}=\frac{1}{2}\bx+\frac{1}{2}\bx^{\perp}$. Take $\bx$ to be
the reference state. According to axiom 3 we can generate the full
set of states by applying all possible transformations to the
reference state. Since the totally mixed state is invariant under
the transformations, the pair of orthogonal states is represented by
a pair of antiparallel vectors $\bx^{\perp}=-\bx$. Consider the set
$\cS_{\mathrm{pure}}=\{~U\bx~|~\forall U\}$ of all pure states
generated by applying all transformations to the reference state. If
one uses the orthogonal representation of the transformations,
$U=S^{-1}OS$, which was introduced above, one maps $ \bx \mapsto
S\bx$ and $U \mapsto O$. Hence, the transformation $U\bx \mapsto
SU\bx= OS\bx$ is norm preserving. We conclude that all pure states
are points on a $d$-dimensional ellipsoid described by $||S\bx||=c$
with $c>0$.

\begin{figure}\centering
\includegraphics[width=8cm]{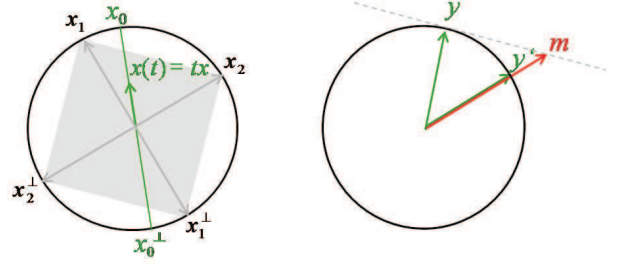}
\caption{(Left) Illustration to the proof that the entire
$d$-dimensional ellipsoid (here represented by a circle; $d=2$)
contains physical states. Consider a line $\bx(t)=t\bx$ through the
origin. A point on the line can be expanded into a set of linearly
independent vectors $\bx_i$ (here $\bx_1$ and $\bx_2$). For
sufficiently small $t$ (i.e. when the line is within the gray
square) the point $\bx(t)$ can be represented as a convex mixture
over $\bx_i$ and their orthogonal vectors $\bx^{\perp}_i$ and thus
is a physical state. According to axiom 1, $\bx(t)$ can be
represented as a convex mixture of two orthogonal pure states
$\bx_0$ and $\bx^{\perp}_0$:
$\bx(t)=t\bx=\alpha\bx_0+(1-\alpha)(-\bx_0)$, where $\bx=\bx_0$ (see
text for details). This implies that every point in the ellipsoid is
a physical state. (Right) Illustration to the proof that in the
orthogonal representation the measurement vector $\bm$ that
identifies the state $\bx$, i.e. for which the probability
$P_{\mathrm{meas}}=\frac{1}{2}(1+\bm^{\mathrm{T}}\by)=1$, is identical to
the state vector, $\bm=\by$. Suppose that $\bm \neq \by$, then
$||\bm|| > 1$, since the state vector is normalized. But then the
same measurement for state $\by'$ parallel to $\bm$ would return a
probability larger than 1, which is nonphysical. Thus
$\bm=\by$.}\label{Picture}
\end{figure}

Now, we want to show that any vector $\bx$ satisfying $||S\bx||=c$
is a physical state and therefore the set of states has to be the
whole ellipsoid. Let $\bx$ be some vector satisfying $||S\bx||=c$
and $\bx(t)=t\bx$ a line trough the origin (totaly mixed state) as
given in Figure ~\ref{Picture} (left). Within the set of pure states
we can always find $d$ linearly independent vectors
$\{\bx_1,\dots,\bx_{d}\}$. For each state $\bx_i$ there is a
corresponding orthogonal state $\bx_i^{\perp}=-\bx_i$ in a set of
states. We can expand a point on the line into a linearly
independent set of vectors: $\bx(t)=t\sum_{i=1}^{d}c_i\bx_i$. For
sufficiently small $t$ we can define a pair of non-negative numbers
$\lambda_i(t)=\frac{1}{2}(\frac{1}{d}+tc_i)$ and
$\lambda^{\perp}_i(t)=\frac{1}{2}(\frac{1}{d}-tc_i)$ with $\sum_i
(\lambda_i(t) +\lambda^{\perp}_i(t))=1$ such that $\bx(t)$ is a
mixture
$\bx(t)=\sum_{i=1}^{d}\lambda_i(t)\bx_i+\lambda_i^{\perp}(t)\bx_i^{\perp}$
and therefore is a physical state. Then, according to axiom 1 there
exists a pair of basis states $\{\bx_0,-\bx_0\}$ such that $\bx(t)$
is a mixture of them
\begin{equation}
\bx(t)=t\bx=\alpha\bx_0+(1-\alpha)(-\bx_0),
\end{equation}
where $\alpha=\frac{1+t}{2}$ and $\bx=\bx_0$. This implies that
$\bx$ is a pure state and therefore all points of the ellipsoid are
physical states.

For every pure state $\bx$, there exists at least one measurement
setting with the outcome $\br$ such that the outcome probability is
one, hence $\br^{\mathrm{T}}\bx=1$. Let us define new coordinates
$\by=\frac{1}{c}S\bx$ and $\bm=cS^{-1T}\br$ in the orthogonal
representation. The set of pure states in the new coordinates is a
$(d-1)$-sphere $\cS^{d-1}=\{\by~|~||\by||=1\}$ of the radius. The
probability rule (\ref{prob rule}) remains unchanged in the new
coordinates:
\begin{equation}
P_{\mathrm{meas}}=\frac{1}{2}(1+\bm^{\mathrm{T}}\by).
\end{equation}
Thus, one has $\bm^{\mathrm{T}}\by=1$. Now, assume that $\bm\neq\by$. Then
$||\bm||>1$ and the vectors $\bm$ and $\by$ span a two-dimensional
plane as illustrated in Figure~\ref{Picture} (right). The set of
pure states within this plane is a unit circle. Choose the pure
state $\by'$ to be parallel to $\bm$. Then the outcome probability
is $P_{\mathrm{measur}}=\frac{1}{2}(1+||\bm||||\by'||)>1$ which is
non-physical, hence $\bm=\by$. Therefore, to each pure state $\by$,
we associate a measurement vector $\bm=\by$ which identifies it.
Equivalently, in the original coordinates, to each $\bx$  we
associate a measurement vector $\br=D\bx$, where
$D=\frac{1}{c^2}S^{\mathrm{T}}S$ is a positive, symmetric matrix. A proof of
this relation for the restricted case of $d=3$ can be found in
Ref.~\cite{Hardy}.

From now one, instead of the measurement vector $\br$ we will use
the pure state $\bx$ which identifies it. When we say that the
measurement along the state $\bx$ is performed we mean the
measurement given by $\br=D\bx$. The measurement setting is given by
a pair of measurement vectors $\br$ and $-\br$. The measured
probability when the state $\bx_1$ is measured along the state
$\bx_2$ follows from formula (\ref{prob rule}):
\begin{equation}
P(\bx_1,\bx_2)=\frac{1}{2}(1+\bx_1^{\mathrm{T}}D\bx_2).
\end{equation}

We can choose orthogonal eigenvectors of the matrix $D$ as the
fiducial set of states (measurements):
\begin{equation}
D\bx_i=a_i\bx_i,
\end{equation}
where $a_i$ are eigenvalues of $D$. Since $\bx_i$ are pure states, they satisfy
$\bx_i^{\mathrm{T}}D\bx_j=\delta_{ij}$. The set of pure states becomes a unit
sphere $\cS^{d-1}=\{\bx~|~||\bx||=1\}$ and the probability formula
is reduced to
\begin{equation}\label{1bit prob}
P(\bx_1,\bx_2)=\frac{1}{2}(1+\bx_1^{\mathrm{T}}\bx_2).
\end{equation}
This corresponds to a choice of a complete set of mutually complementary measurements
(i.e. mutually unbiased basis sets) for the fiducial measurements. The states identifying  outcomes of complementary measurements satisfy $P(\bx_i,\bx_j)=\frac{1}{2}$ for $i\neq j$. Two observables are
said to be mutually complementary if complete certainty about one of
the observables (one of two outcomes occurs with  probability one)
precludes any knowledge about the others (the probability for both
outcomes is $1/2$). Given some state $\bx$, the $i$-th fiducial
measurement returns probability $p_i=\frac{1}{2}(1+x_i)$. Therefore,
$x_i$ is a mean value of a dichotomic observable
$\bb_i=+1\bx_i-1\bx_i^{\perp}$ with two possible outcomes
$b_i=\pm1$.

A theory in which the state space of the generalized bit is
represented by a $(d-1)$-sphere has $d$ mutually complementary
observables. This is a characteristic feature of the theories and
they can be ordered according to their number. For example,
classical physics has no complementary observables, real quantum
mechanics has two, complex (standard) quantum mechanics has three
(e.g. the spin projections of a spin-1/2 system along three
orthogonal directions) and the one based on quaternions has five
mutually complementary observables. Note that higher-order theories
of a single generalized bit are such that the qubit
theory can be embedded in them in the same way in which classical
theory of a bit can be embedded in qubit theory itself.

Higher-order theories can have even better information processing
capacity than quantum theory. For example, the computational
abilities of the theories with $d=2^r$ and $r\in N$ in solving the
Deutsch-Josza type of problems increases with the number of mutually
complementary measurements~\cite{Paterek}. It is likely that the
larger this number is the larger the error rate would be in secret
key distribution in these theories, in a similar manner in which the
6-state is advantageous over the 4-state protocol in (standard)
quantum mechanics. In the first case one uses all three mutually
complementary observables and in the second one only two of them.
(See Ref.~\cite{Barnum3} for a review on characterizing generalized
probabilistic theories in terms of their information-processing power
and Ref.~\cite{Abramsky} for investigating the same question in much
more general framework of compact closed categories.)

A final remark on higher-order theories is of more speculative
nature. In various approaches to quantum theory of gravity one
predicts at the Planck scale the dimension of space-time to be
different from $3+1$~\cite{Loll}. If one considers directional
degrees of freedom (spin), then the $d-1$-sphere (Bloch sphere)
might be interpreted as the state space of a spin system embedded in
real (ordinary) space of dimension $d$, in general different than 3
which is the special case of quantum theory.

The reversible transformation $R$ preserves the purity of state
$||R\bx||=||x||$ and therefore $R$ is an orthogonal matrix.
We have shown that the state space is the full $(d-1)$-sphere. According to axiom 3 the set of transformations must be rich enough to generate the full sphere. If $d=1$ (classical bit), the
group of transformations is discrete and contains only the identity and
the bit-flip. If $d>1$, the group is continuous and is some subgroup of the orthogonal group
$\mathrm{O}(d)$. Every orthogonal matrix has determinant either 1 or
-1. The orthogonal matrices with determinant 1 form a normal
subgroup of $\mathrm{O}(d)$, known as the special orthogonal group
$\mathrm{SO}(d)$. The group $\mathrm{O}(d)$ has two connected
components: the identity component which is the $\mathrm{SO}(d)$ group, and the component formed by orthogonal
matrices with determinant -1. Since every two points on the
$(d-1)$-sphere are connected by some transformation, the group of
transformations is at least the $\mathrm{SO}(d)$ group. If we
include even a single transformation with determinant -1, the set of
transformations becomes the entire $\mathrm{O}(d)$ group. (Later we will show that only some $d$ are in agreement with our three axioms and for these $d$'s the set of physical transformations will be shown to be the $\mathrm{SO}(d)$ group).

\section{Composite system and the notion of locality}

We now introduce a description of composite systems. We assume that
when one combines two systems of dimension $L_1$ and $L_2$ into a
composite one, one obtains a system of dimension $L_1L_2$. Consider
a composite system consisting of two geneneralized bits and choose a
set of $d$ complementary measurements on each subsystem as fiducial
measurements. According to axiom 2 the state of the composite system
is completely determined by a set of real parameters obtainable from
local measurements on the two generalized bits and their
correlations. We obtain $2d$ independent real parameters from the
set of local fiducial measurements and additional $d^2$ parameters
from correlations between them. This gives  altogether
$d^2+2d=(d+1)^2-1$ parameters. They are the components $x_i$, $y_i$,
$i\in \{1,...,d\}$, of the local Bloch vectors and $T_{ij}$ of the
correlation tensor:
\begin{eqnarray}
x_i=p^{(i)}(A=1)-p^{(i)}(A=-1),\\
y_j=p^{(j)}(B=1)-p^{(j)}(B=-1),\\
T_{ij}=p^{(ij)}(AB=1) -p^{(ij)}(AB=-1).
\end{eqnarray}
Here, for example, $p^{(i)}(A=1)$ is the probability to obtain
outcome $A=1$ when the $i$-th measurement is performed on the first
subsystem and $p^{(ij)}(AB=1)$ is the joint probability to obtain
correlated results (i.e. either $A=B=+1$ or $A=B=-1$) when the
$i$-th measurement is performed on the first subsystem and the
$j$-th measurement on the second one.

Note that axiom 2 ``The state of a composite system is completely
determined by local measurements on its subsystems and their
correlations'' is formulated in a way that the non-signaling
condition is implicitly assumed to hold. This is because it is
sufficient to speak about ``local measurements'' alone without
specifying the choice of measurement setting on the other,
potentially distant, subsystem. Therefore, $x_i$ does not depend on
$j$, and $y_j$ does not depend on $i$.

We represent a state by the triple $\psi=(\bx,\by,T)$, where $\bx$
and $\by$ are the local Bloch vectors and $T$ is a $d\times d$ real
matrix representing the correlation tensor. The product (separable)
state is represented by $\psi_p=(\bx,\by,T)$, where
$T=\bx\by^{\mathrm{T}}$ is of product form, because the correlations are
just products of the components of the local Bloch vectors. We call
the pure state \emph{entangled} if it is not a product state.

The measured probability is a linear function of the state $\psi$.
If we prepare totaly mixed states of the subsystems $(0,0,0)$, the
probability for any outcome of an arbitrary measurement will be
$1/4$. Therefore, the outcome probability can be written as:
\begin{equation}\label{2bit prob_rule}
P_{\mathrm{measur}}=\frac{1}{4}(1+(r,\psi)),
\end{equation}
where $r=(\br_1,\br_2,K)$ is a measurement vector associated to
the observed outcome and $(...,...)$ denotes the scalar product:
\begin{equation}(r,\psi)=\br_1^{\mathrm{T}}\bx+\br_2^{\mathrm{T}}\by+\mathrm{Tr}(K^{\mathrm{T}}T).
\end{equation}

Now, assume that $r=(\br_1,\br_2,K)$ is associated to the outcome
which is identified by some product state
$\psi_p=(\bx_0,\by_0,T_0)$. If we preform a measurement on the
arbitrary product state $\psi=(\bx,\by,T)$, the outcome probability
has to factorize into the product of the local outcome probabilities
of the form (\ref{1bit prob}):
\begin{eqnarray}
P_{\mathrm{measur}}&=&\frac{1}{4}(1+\br_1^{\mathrm{T}}\bx+\br_2^{\mathrm{T}}\by+\bx^{\mathrm{T}}K\by)\\
&=&P_1(\bx_0,\bx)P_2(\by_0,\by)\\
&=&\frac{1}{2}(1+\bx_0^{\mathrm{T}}\bx)\frac{1}{2}(1+\by_0^{\mathrm{T}}\by)\\
&=&\frac{1}{4}(1+\bx_0^{\mathrm{T}}\bx+\by_0^{T}\by+\bx^{\mathrm{T}}\bx_0\by_0^{\mathrm{T}}\by),
\end{eqnarray}
which holds for all $\bx,\by$. Therefore we have $r=\psi_p$. For
each product state $\psi_p$ there is a unique outcome $r=\psi_p$
which identifies it. We will later show that correspondence
$r=\psi$ holds for {\em all} pure states $\psi$.

If we preform local transformations $R_1$ and $R_2$ on the
subsystems, the global state $\psi=(\bx,\by,T)$ is transformed to
\begin{equation}\label{local transform}
(R_1,R_2)\psi=(R_1\bx,R_2\by,R_1TR_2^{\mathrm{T}}).
\end{equation}
$T$ is a real matrix and we can find its singular value
decomposition $\mathrm{diag}[t_1,\dots,t_d]=R_1TR_2^{\mathrm{T}}$, where
$R_1,R_2$ are orthogonal matrices which can be chosen to have
determinant 1. Therefore, we can choose the local bases such that
correlation tensor $T$ is a diagonal matrix:
\begin{equation}\label{Schmidt form}
(R_1,R_2)(\bx,\by,T)=(R_1\bx,R_2\by,\mathrm{diag}[t_1,\dots,t_d]).
\end{equation}
The last expression is called \emph{Schmidt decomposition} of the
state.

The local Bloch vectors satisfy $||\bx||,||\by||\leq1$ which implies
a bound on the correlation $||T||\geq 1$ for all pure states. The
following lemma
 identifies a simple entanglement witness for pure states. The proof of this and all subsequent lemmas is given in the Appendix.
\begin{Lemma}The lower bound $||T||=1$ is saturated, if and only if
the state is a product state $T=\bx\by^{\mathrm{T}}$.
\end{Lemma}

Recall that for every transformation $U$ we can find its orthogonal
representation $U=SOS^{-1}$ (the Schur-Auerbach lemma), where $S$ is
an invertible matrix and $O^{\mathrm{T}}O=\openone$. The matrix $S$ is
characteristic of the representation and should be the same for all
transformations $U$. If we choose some local transformation
$U=(R_1,R_2)$, $U$ will be orthogonal and thus we can choose to set
$S=\openone$. The representation of transformations is orthogonal,
therefore they are norm preserving. By applying simultaneously all
(local and non-local) transformations $U$ to some product state (the
reference state) $\psi$ and to the measurement vector which identifies
it, $r=\psi$, we generate the set of all pure states and
corresponding measurement vectors. Since we have
$1=P(r=\psi,\psi)=P(Ur,U\psi)$, correspondence $r=\psi$ holds for
any pure state $\psi$. Instead of the measurement vector $r$ in
formula (\ref{2bit prob_rule}) we use the pure state which
identifies it. If the state $\psi_1=(\bx_1,\by_1,T_1)$ is prepared
and measurement along the state $\psi_2=(\bx_2,\by_2,T_2)$ is
performed, the measured probability is given by
\begin{eqnarray}
P_{12}(\psi_1,\psi_2)=\frac{1}{4}(1+\bx_1^{\mathrm{T}}\bx_2+\by_1^{\mathrm{T}}\by_2+\mathrm{Tr}(T_1^{\mathrm{T}}T_2)).
\end{eqnarray}
The set of pure states obeys $P_{12}(\psi,\psi)=1$. We can define
the normalization condition for pure states
$P_{12}(\psi,\psi)=\frac{1}{4}(1+||\bx||^2+||\by||^2+||T||^2)=1$
where $||T||^2=\mathrm{Tr}(T^{\mathrm{T}}T)$. Therefore we have:
\begin{equation}\label{norm}
||\bx||^2+||\by||^2+||T||^2=3,
\end{equation}
for all pure states.

An interesting observation can be made here. Although seemingly
axiom 2 does not imply any strong prior restrictions to $d$, we
surprisingly have obtained the explicit number 3 in the
normalization condition~(\ref{norm}). As we will see soon this
relation will play an important role in deriving $d=3$ as the only
non-classical solution consistent with the axioms.

\section{The Main Proofs}

We will now show that only classical probability theory and quantum
theory are in agreement with the three axioms.

\subsection{Ruling out the $d$ even case}

Let us assume the total inversion $E\bx=-\bx$ being a physical
transformation. Let $\psi=(\bx,\by,T)$ be a pure state of composite
system. We apply total inversion to one of the subsystems and obtain
the state $\psi'=(E,\openone)(\bx,\by,T)=(-\bx,\by,-T)$. The
probability
\begin{eqnarray}
P_{12}(\psi,\psi')&=&\frac{1}{4}(1-||\bx||^2+||\by||^2-||T||^2)\\
&=&\frac{1}{2}(||\by||^2-1)
\end{eqnarray}
has to be nonnegative and therefore we have $||\by||=1$. Similarly,
we apply $(\openone,E)$ to $\psi$ and obtain $||\bx||=1$. Since the
local vectors are of the unit norm we have $||T||=1$ and thus,
according to lemma 1, the state $\psi$ is a product state. We
conclude that no entangled states can exist if $E$ is a physical
transformation. As we will soon see, according to axiom 1 entangled
states must exist. Thus, $E$ cannot represent a physical
transformation. We will now show that this implies that $d$ has to be odd.  Recall that the set of transformations is at least the $\mathrm{SO}(d)$ group. $d$ cannot be even since $E$ would have unit determinant and would belong to $\mathrm{SO}(d)$. $d$ has to be odd in which case $E$ has determinant -1. The set of physical transformations is the $\mathrm{SO}(d)$ group.

\subsection{Ruling out the $d>3$ case.}

Let us define one basis set of two generalized bit product states:
\begin{eqnarray}
\psi_1&=&(\be_1,\be_1,T_0=\be_1\be_1^{\mathrm{T}})\\
\psi_2&=&(-\be_1,-\be_1,T_0)\\
\psi_3&=&(-\be_1,\be_1,-T_0)\\
\psi_4&=&(\be_1,-\be_1,-T_0)
\end{eqnarray}
with $\be_1=(1,0,\dots,0)^{\mathrm{T}}$. Now, we define two subspaces $S_{12}$
and $S_{34}$ spanned by the states $\psi_1,\psi_2$ and
$\psi_3,\psi_4$, respectively. Axiom 1 states that these
two subspaces behave like one-bit spaces, therefore they are
isomorphic to the $(d-1)$-sphere $S_{12}\cong S_{34}\cong
\cS^{d-1}$. The state $\psi$ belongs to $S_{12}$ if and only if the following
holds:
\begin{equation}\label{12}
P_{12}(\psi,\psi_1)+P_{12}(\psi,\psi_2)=1.
\end{equation}
Since the $\psi_1,\dots,\psi_4$ form a complete basis set, we have
\begin{equation}\label{34}
P_{12}(\psi,\psi_3)=0,~~P_{12}(\psi,\psi_4)=0.
\end{equation}
A similar reasoning holds for states belonging to the $S_{34}$
subspace. Since the states $\psi\in S_{12}$ and $\psi'\in S_{34}$
are perfectly distinguishable in a single shot experiment, we have
$P_{12}(\psi,\psi')=0$. Therefore, $S_{12}$ and $S_{34}$ are
orthogonal subspaces.

Axiom 1 requires the existence of entangled states as it is apparent
from the following Lemma 2.
\begin{Lemma} The only product states belonging to $S_{12}$ are $\psi_1$ and $\psi_2$.
\end{Lemma}

We define a local mapping between orthogonal subspaces $S_{12}$ and
$S_{34}$. Let the state $\psi=(\bx,\by,T)\in S_{12}$, with
$\bx=(x_1,x_2,\dots,x_d)^{\mathrm{T}}$ and $\by=(y_1,y_2,\dots,y_d)^{\mathrm{T}}$.
Consider the one-bit transformation $R$ with the property
$R\be_1=-\be_1$. The local transformation of this type maps the
state from $S_{12}$ to $S_{34}$ as shown by the following lemma:
\begin{Lemma}\label{sub flip}If the state $\psi\in S_{12}$, then $\psi'=(R,\openone)\psi\in S_{34}$ and $\psi''=(\openone,R)\psi\in
S_{34}$.
\end{Lemma}

Let us define $\bT_i^{(x)}=(T_{i1},\dots,T_{id})$ and
$\bT_i^{(y)}=(T_{1i},\dots,T_{di})^{\mathrm{T}}$. The correlation tensor can be
rewritten in two different ways:
\begin{equation}
T=\left(
    \begin{array}{c}
      \bT_1^{(x)} \\
      \bT_2^{(x)} \\
      \vdots \\
      \bT_d^{(x)} \\
    \end{array}
  \right)~~\rm{or}~~T=\left(
                  \begin{array}{cccc}
                    \bT_1^{(y)} & \bT_2^{(y)} & \dots & \bT_d^{(y)} \\
                  \end{array}
                \right).
\end{equation}

Consider now the case $d>3$. We define local transformations $R_i$
flipping the first and $i$-th coordinate and $R_{jkl}$ flipping the
first and $j$-th, $k$-th, and $l$-th coordinate with $j\neq k\neq
l\neq1$. Let $\psi=(\bx,\by,(\bT_1^{(x)},\dots,\bT_d^{(x)})^{\mathrm{T}})$
belong to $S_{12}$. According to Lemma 2, the states
$\psi_i=(R_i,\openone)\psi$ and $\psi_{jkl}=(R_{jkl},\openone)\psi$
belong to $S_{34}$, therefore $P_{12}(\psi,\psi_i)=0$ and
$P_{12}(\psi,\psi_{jkl})=0$. We have:
\begin{eqnarray}
0&=&P_{12}(\psi,\psi_i)\\
&&1-x_1^2+x_2^2+\dots-x_i^2+\dots+x_d^2+||\by||^2\\
&&-||\bT_1^{(x)}||^2+||\bT_2^{(x)}||^2+\dots-||\bT_i^{(x)}||^2+\dots+||\bT_d^{(x)}||^2\\
&=&1-2x_1^2-2x_i^2-2||\bT_1^{(x)}||^2-2||\bT_i^{(x)}||^2+||\bx||^2+||\by||^2+||T||^2 \nonumber \\
&=&2(2-x_1^2-x_i^2-||\bT_1^{(x)}||^2-||\bT_i^{(x)}||^2).
\end{eqnarray}
Similarly, we expand $P_{12}(\psi,\psi_{jkl})=0$ and together with
the last equation we obtain:
\begin{eqnarray}\label{cond1}
&&x_1^2+x_i^2+||\bT_1^{(x)}||^2+||\bT_i^{(x)}||^2=2\\\label{cond2}
&&x_1^2+x_j^2+x_k^2+x_l^2+||\bT_1^{(x)}||^2+||\bT_j^{(x)}||^2+||\bT_k^{(x)}||^2+||\bT_l^{(x)}||^2=2. \nonumber
\end{eqnarray}
Since this has to hold for all $i,j,k,l$ we have:
\begin{eqnarray}
&&x_2=x_3=\dots=x_d=0\\
&&\bT_2^{(x)}=\bT_3^{(x)}=\dots=\bT_d^{(x)}=0.
\end{eqnarray}
We repeat this kind of reasoning for the transformations
$(\openone,R_i)$ and $(\openone,R_{jkl})$ and obtain:
\begin{eqnarray}\label{cond3}
&&y_1^2+y_i^2+||\bT_1^{(y)}||^2+||\bT_i^{(y)}||^2=2\\\label{cond4}
&&y_1^2+y_j^2+y_k^2+y_l^2+||\bT_1^{(y)}||^2+||\bT_j^{(y)}||^2+||\bT_k^{(y)}||^2+||\bT_l^{(y)}||^2=2. \nonumber
\end{eqnarray}
Therefore, we have
\begin{eqnarray}
&&y_2=y_3=\dots=y_d=0\\
&&\bT_2^{(y)}=\bT_3^{(y)}=\dots=\bT_d^{(y)}=0.
\end{eqnarray}
The only non-zero element of the correlation tensor is $T_{11}$ and
it has to be exactly 1, since $||T||\geq1$. This implies that $\psi$
is a product state, furthermore $\psi=\psi_1$ or $\psi=\psi_2$.

This concludes our proof that only the cases $d=1$ and $d=3$ are in
agreement with our three axioms. To distinguish between the two
cases, one can invoke the continuity axiom (3') and proceed as in
the reconstruction given by Hardy~\cite{Hardy}.

\section{``Two'' quantum mechanics}

We now obtain two solutions for the theory of a composite system
consisting of two bits in the case when $d=3$. One of them
corresponds to the standard quantum theory of two qubits, the other
one to its ``mirror'' version in which the states are obtained from
the ones from the standard theory by partial transposition. Both
solutions are regular as far as one considers composite systems of
two bits, but the ``mirror'' one cannot be consistently constructed
already for systems of three bits.

Two conditions (\ref{12}) and (\ref{34}) put the constraint to the form of $\psi$:
\begin{eqnarray}
x_1=-y_1,~~~~T_{11}=1.
\end{eqnarray}
The subspace $S_{12}$ is isomorphic to the sphere $\cS^2$. Let us
choose $\psi$ complementary to the one bit basis $\{\psi_1,\psi_2\}$
in $S_{12}$. We have $P_{12}(\psi,\psi_1)=P_{12}(\psi,\psi_2)=1/2$
and thus $x_1=y_1=0$. For simplicity we write $\psi$ in the form:
\begin{equation}\label{state12}
\psi=\left(\left(
        \begin{array}{c}
          0 \\
          \bx \\
        \end{array}
      \right),\left(
        \begin{array}{c}
          0 \\
          \by \\
        \end{array}
      \right),
      \left(
        \begin{array}{cc}
          1 & \bT_y^{\mathrm{T}} \\
          \bT_x & T \\
        \end{array}
      \right)
\right),
\end{equation}
with $\bx=(x_2,x_3)^{\mathrm{T}}$, $\by=(y_2,y_3)^{\mathrm{T}}$,
$\bT_y=(T_{12},T_{13})^{\mathrm{T}}$, $\bT_x=(T_{21},T_{31})^{\mathrm{T}}$ and $T=\left(
\begin{array}{cc}
T_{22} & T_{23} \\
T_{32} & T_{33} \\
\end{array}
\right) $.

Let $R(\phi)$ be a rotation around the $\be_1$ axis. This
transformation keeps $S_{12}$ invariant. Now, we show that the state
$\psi$ as given by equation (\ref{state12}) cannot be invariant
under local transformation $(\openone,R(\phi))$. To prove this by
{\em reductio ad absurdum} suppose the opposite, i.e. that
$(\openone,R(\phi))\psi=\psi$. We have three conditions
\begin{equation}
R(\phi)\by=\by,~~~\bT_y^{\mathrm{T}}R^{\mathrm{T}}(\phi)=\bT_y^{\mathrm{T}},~~~TR^{\mathrm{T}}(\phi)=T,
\end{equation}
which implies $\by=0$, $\bT_y^{\mathrm{T}}=0$ and $T=0$ thus
\begin{equation}
\psi=\left(\left(
        \begin{array}{c}
          0 \\
          x_2 \\
          x_3 \\
        \end{array}
      \right),\left(
        \begin{array}{c}
          0 \\
          0 \\
          0 \\
        \end{array}
      \right),
      \left(
        \begin{array}{ccc}
          1 & 0 & 0 \\
          T_2 & 0 & 0 \\
          T_3 & 0 & 0 \\
        \end{array}
      \right)
\right).
\end{equation}
According to equations (\ref{cond1}) and (\ref{cond3}) we can easily
check that $||\bx||=1$, and thus $\psi$ is locally equivalent to the
state:
\begin{equation}
\psi'=\left(\left(
        \begin{array}{c}
          0 \\
          0 \\
          1 \\
        \end{array}
      \right),\left(
        \begin{array}{c}
          0 \\
          0 \\
          0 \\
        \end{array}
      \right),
      \left(
        \begin{array}{ccc}
          1 & 0 & 0 \\
          T'_2 & 0 & 0 \\
          T'_3 & 0 & 0 \\
        \end{array}
      \right)
\right).
\end{equation}
Let $\chi_{1}=(-\be_3,\be_1,-\be_3\be_1^{\mathrm{T}})$ and
$\chi_{2}=(-\be_3,-\be_1,\be_3\be_1^{\mathrm{T}})$. The two conditions
$P(\psi',\chi_1)\geq0$ and $P(\psi',\chi_2)\geq0$ become
\begin{eqnarray}
&&\frac{1}{4}(1-1-T'_3)=-\frac{1}{4}T'_3\geq0\\
&&\frac{1}{4}(1-1+T'_3)=\frac{1}{4}T'_3\geq0
\end{eqnarray}
and thus $T'_3=0$. The normalization condition~({\ref{norm}) gives $T'_2=\pm 1$. The state $\psi'$ is
not physical. This can be seen when one performs the rotation $(R,\openone)$ where 
\begin{equation}
R=\left(
    \begin{array}{ccc}
      \tfrac{1}{\sqrt2} & -\tfrac{1}{\sqrt2} & 0 \\
      \tfrac{1}{\sqrt2} & \tfrac{1}{\sqrt2} & 0 \\
      0 & 0 & 1 \\
    \end{array}
  \right).
\end{equation}
The transformed correlation tensor has a component $\sqrt{2}$ which is non-physical. 
Therefore, the transformation $(\openone,R(\phi))\psi$
draws a full circle of pure states in a plane orthogonal to $\psi_1$
within the subspace $S_{12}$. Similarly, the transformation
$(R(\phi),\openone)$ draws the same set of pure states when applied
to $\psi$. Hence, for every transformation $(\openone,R(\phi_1))$
there exists a transformation $(R(\phi_2),\openone)$ such that
$(\openone,R(\phi_1))\psi=(R(\phi_2),\openone)\psi$. This gives us a
set of conditions:
\begin{eqnarray}
R(\phi_2)\bx&=&\bx\\
R(\phi_1)\by&=&\by\\
R(\phi_2)\bT_x&=&\bT_x\\
\bT_y^{\mathrm{T}}R^{\mathrm{T}}(\phi_1)&=&\bT_y^{\mathrm{T}}\\
R(\phi_2)T&=&TR^{\mathrm{T}}(\phi_1),
\end{eqnarray}
which are fulfilled if $\bx=\by=\bT_x=\bT_y=0$ and
$T=\mathrm{diag}[T_1,T_2]$. Equation (\ref{cond1}) gives
$T_2^2=T_3^2=1$ and we finally end up with two different solutions:
\begin{equation}
\psi_{\mathrm{QM}}=(0,0,\mathrm{diag}[1,-1,1])~~\vee~~\psi_{\mathrm{MQM}}=(0,0,\mathrm{diag}[1,1,1]).
\end{equation}

\begin{figure}\centering
\includegraphics[width=8.7cm]{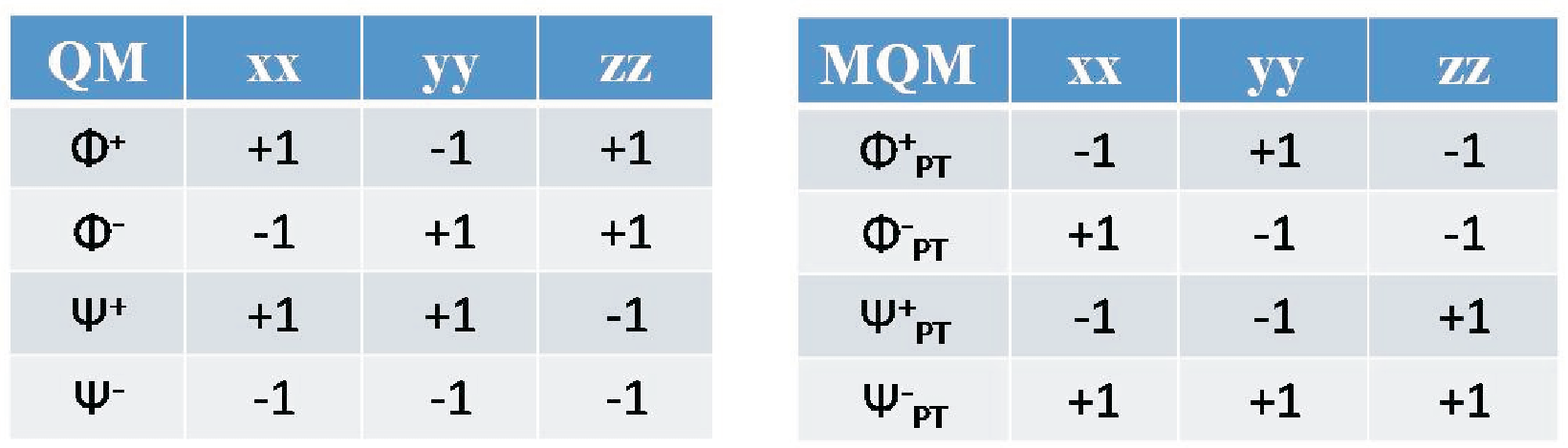}
\caption{Correlations between results obtained in measurements of
two bits in a maximal entangled (Bell's) state in standard quantum
mechanics (Left) and ``mirror quantum mechanics'' (Right) along $x$,
$y$ and $z$ directions. Why do we never see correlations as given in
the table on the right? The opposite sign of correlations on the right and on the left is not a matter of convention or labeling of outcomes. If one can transport the two bits parallel to the same detector, one can distinguish operationally between the two types of correlations~\cite{Rob}.} \label{tables}
\end{figure}

The first ``M'' in $\psi_{\mathrm{MQM}}$ stands for ``mirror''. The
two solutions are incompatible and cannot coexist within the same
theory. The first solution corresponds to the triplet state
$\phi^{+}$ of ordinary quantum mechanics. The second solution is a
totally invariant state and has a negative overlap with, for
example, the singlet state $\psi^-$ for which
$T=\mathrm{diag}[-1,-1,-1]$. That is, if the system were prepared in
one of the two states and the other one were measured, the
probability would be negative. Nevertheless, both solutions are
regular at the level of two bits. The first belongs to ordinary
quantum mechanics with the singlet in the ``antiparallel'' subspace
$S_{34}$ and the second solution is ``the singlet state in the
parallel subspace'' $S_{12}$. We will show that one can build the
full state space, transformations and measurements in both cases.
The states from one quantum mechanics can be obtained from the other
by partial transposition
$\psi_{\mathrm{QM}}^{\mathrm{PT}}=\psi_{\mathrm{MQM}}$. In
particular, the four maximal entangled states (Bell states) from
``mirror  quantum mechanics'' have correlations of the opposite
sign of those from the standard quantum mechanics (see Figure 4).

Now we show that the theory with ``mirror states'' is physically
inconsistent when applied to composite system of three bits. Let us
first derive the full set of states and transformations for two
qubits in standard quantum mechanics. We have seen that the state
$\psi_{\mathrm{QM}}$ belongs to the subspace $S_{12}$, and
furthermore, that it is complementary (within $S_{12}$) to the
product states $\psi_1$ and $\psi_2$. The totally mixed state within
the $S_{12}$ subspace is
$E_{12}=\frac{1}{2}\psi_1+\frac{1}{2}\psi_2$. The states $\psi_1$
and $\psi_{\mathrm{QM}}$ span one two-dimensional plane, and the set
of pure states within this plane is a circle:
\begin{eqnarray}\label{schmidt states}
\psi(x)&=&E_{12}+\cos x~(\psi_1-E_{12})+\sin x~(\psi_{\mathrm{QM}}-E_{12})\\
&=&(\cos x~\be_1,\cos x~\be_1,\mathrm{diag}[1,-\sin x,\sin x]).
\end{eqnarray}

We can apply a complete set of local transformations to the set
$\psi(x)$ to obtain the set of all pure two-qubit states. Let us
represent a pure state $\psi=(\bx,\by,T)$ by the $4\times4$
Hermitian matrix $\rho$:
\begin{equation}
\rho=\frac{1}{4}(\openone\otimes\openone+\sum_{i=1}^3x_i\sigma_i\otimes\openone+\sum_{i=1}^3y_i\openone\otimes\sigma_i+
\sum_{i,j=1}^3T_{ij}\sigma_i\otimes\sigma_j),
\end{equation}
where $\sigma_i$, $i \in \{1,2,3\}$, are the three Pauli matrices.
It is easy to show that the set of states (\ref{schmidt states})
corresponds to the set of one-dimensional projectors
$\ket{\psi(x)}\bra{\psi(x)}$, where
$\ket{\psi(x)}=\cos{\frac{x}{2}}\ket{00}+\sin{\frac{x}{2}}\ket{11}$.
The action of local transformations $(R_1,R_2)\psi$ corresponds to
local unitary transformation $U_1\otimes
U_2\ket{\psi}\bra{\psi}U_1^{\dagger}\otimes U_2^{\dagger}$, where
the correspondence between $U$ and $R$ is given by the isomorphism
between the groups $\mathrm{SU}(2)$ and $\mathrm{SO}(3)$:
\begin{equation}
U\rho
U^{\dagger}=\frac{1}{2}\left(\openone+\sum_{i=1}^3\left(\sum_{j=1}^3R_{ij}x_j\right)\sigma_i\right).
\end{equation}
Here $R_{ij}=\mathrm{Tr}(\sigma_i U\sigma_j U^{\dagger})$ and
$x_i=\mathrm{Tr}\sigma_i\rho$. When we apply a complete set of local
transformations to the states $\ket{\psi(x)}$ we obtain the whole
set of pure states for two qubits. The group of transformations is
the set of unitary transformations $\mathrm{SU}(4)$.

The set of states from ``mirror quantum mechanics'' can be obtained
by applying partial transposition to the set of quantum states.
Formally, partial transposition with respect to subsystem $1$ is
defined by action on a set of product operators:
\begin{equation}
\mathrm{PT}_1(\rho_1\otimes\rho_2)=\rho_1^{\mathrm{T}}\otimes\rho_2.
\end{equation}
where $\rho_1$ and $\rho_2$ are arbitrary operators. Similarly, we
can define the partial transposition with respect to subsystem $2$,
$\mathrm{PT}_2$. To each unitary transformation $U$ in quantum
mechanics we define the corresponding transformation in ``mirror
mechanics'', e.g. with respect to subsystem 1:
$\mathrm{PT}_1U\mathrm{PT}_1$. Therefore, the set of transformations
is a conjugate group
$\mathrm{PT}_1\mathrm{SU}(4)\mathrm{PT}_1:=\{\mathrm{PT}_1U\mathrm{PT}_1~|~U\in\mathrm{SU}(4)\}$.
Note that we could equally have chosen to apply partial
transposition with respect to subsystem 2, and would obtain the same
set of states. In fact, one can show that
$\mathrm{PT}_1U\mathrm{PT}_1=\mathrm{PT}_2U^{\ast}\mathrm{PT}_2$,
where $U^{\ast}$ is a conjugate unitary transformation (see Lemma 4
in the Appendix). Therefore, the two conjugate groups are the same
$\mathrm{PT}_1\mathrm{SU}(4)\mathrm{PT}_1=\mathrm{PT}_2\mathrm{SU}(4)\mathrm{PT}_2$.
We can generate the set of ``mirror states'' by applying all the
transformations $\mathrm{PT}U\mathrm{PT}$ to some product state,
regardless of which particular partial transposition is used.

Now, we show that ``mirror mechanics'' cannot be consistently
extended to composite systems consisting of three bits. Let
$\psi_p=(\bx,\by,\bz,T_{12},T_{13},T_{23},T_{123})$ be some product
state of three bits, where $\bx$, $\by$ and $\bz$ are local Bloch
vectors, $T_{12}$, $T_{13}$, $T_{23}$ and $T_{123}$ are two- and
three-body correlation tensors, respectively. We can apply the
transformations $\mathrm{PT}U_{ij}\mathrm{PT}$ to a composite system
of $i$ and $j$, and we are free to choose with respect to which
subsystem ($i$ or $j$) to take the partial transposition.
Furthermore, we can combine transformations in $12$ and $13$
subsystems such that the resulting state is genuine three-partite
entangled, and we can choose to partially transpose subsystem 2 in
both cases. We obtain the transformation
\begin{eqnarray}
U_{123}&=&\mathrm{PT}_2U_{12}\mathrm{PT}_2\mathrm{PT}_2U_{23}\mathrm{PT}_2\\
&=&\mathrm{PT}_2U_{12}U_{23}\mathrm{PT}_2.
\end{eqnarray}

When we apply $U_{123}$ to $\psi_p$ we obtain the state
$\mathrm{PT}_2U_{12}U_{23}\phi_p$, where
$\phi_p=\mathrm{PT}_2\psi_p$ is again some product state. The state
$U_{12}U_{23}\phi_p$ is a quantum three qubit state. Since states
$\psi_p$ and $\phi_p$ are product states and do belong to standard
quantum states, we can use the formalism of quantum mechanics and
denote them as $\ket{\psi_p}$ and $\ket{\phi_p}$. Furthermore, since
the state $\ket{\psi_p}$ is an arbitrary product state, without loss
of generality we set $\ket{\phi_p}=\ket{0}\ket{0}\ket{0}$. We can
choose $U_{12}$ and $U_{23}$ such that:
\begin{eqnarray}
U_{12}\ket{0}\ket{0}&=&\ket{0}\ket{0}\\
U_{12}\ket{0}\ket{1}&=&\frac{1}{\sqrt2}(\ket{0}\ket{1}+\ket{1}\ket{0})\\
U_{23}\ket{0}\ket{0}&=&\frac{1}{\sqrt3}\ket{0}\ket{1}+\sqrt{\frac{2}{3}}\ket{1}\ket{0}.
\end{eqnarray}
This way we can generate the $W$-state
\begin{eqnarray}
\ket{W}&=&U_{12}U_{23}\ket{0}\ket{0}\ket{0}\\
&=&\frac{1}{\sqrt{3}}(\ket{0}\ket{0}\ket{1}+\ket{0}\ket{1}\ket{0}+\ket{1}\ket{0}\ket{0}).
\end{eqnarray}
When we apply partial transposition with respect to subsystem 2, we
obtain the corresponding ``mirror W-state'' which we denote as
$W_{\mathrm{M}}$-state, $W_{\mathrm{M}}=\mathrm{PT}_2W$. The local
Bloch vectors and two-body correlation tensors for the $W$ state are
\begin{eqnarray}
&&\bx=\by=\bz=(0,0,\tfrac{1}{3})^{\mathrm{T}},\\
&&T_{12}=T_{13}=T_{23}=\mathrm{diag}[\tfrac{2}{3},\tfrac{2}{3},-\tfrac{1}{3}],
\end{eqnarray}
where $|0\rangle$ corresponds to result +1. Consequently, the local Bloch vectors and the correlation tensor for
$W_{\mathrm{M}}$-state are
\begin{eqnarray}
&&\bx=\by=\bz=(0,0,\tfrac{1}{3})^{\mathrm{T}},\\
&&T_{12}=T_{23}=\mathrm{diag}[\tfrac{2}{3},-\tfrac{2}{3},-\tfrac{1}{3}],\\
&&T_{13}=\mathrm{diag}[\tfrac{2}{3},\tfrac{2}{3},-\tfrac{1}{3}].
\end{eqnarray}

The asymmetry in the signs of correlations in the tensors
$T_{12},T_{23}$ and $T_{13}$ leads to inconsistencies because they
define three different reduced states
$\psi_{ij}=(\bx_i,\bx_j,T_{ij})$, $ij \in \{12,23,13\}$, which
cannot coexist within a single theory. The states $\psi_{12}$ and
$\psi_{23}$ belong to ``mirror quantum mechanics'', while the state
$\psi_{13}$ belongs to ordinary quantum mechanics. To see this, take
the state $\psi=(0,0,\mathrm{diag}[-1,-1,1])$ which is locally
equivalent to state $\psi_{\mathrm{MQM}}=(0,0,\openone)$. The
overlap (measured probability) between the states $\psi_{13}$ and
$\psi$ is negative
\begin{equation}
P(\psi,\psi_{13})=\frac{1}{4}(1-\frac{2}{3}-\frac{2}{3}-\frac{1}{3})=-\frac{1}{6}.
\end{equation}

We conclude that ``mirror quantum mechanics'' -- while being a
perfectly regular solution for a theory of two bits -- cannot be
consistently extended to also describe systems consisting of many
bits. This also answers the question why we find in nature only four
types of correlations as given in the table (Figure 4) on the left,
rather than all eight logically possible ones.

\section{Higher-dimensional Systems and State Up-date Rule in Measurement}

Having obtained $d=3$ for a two-dimensional system we have derived
quantum theory of this system. We have also reconstructed quantum
mechanics of a composite system consisting of two qubits. Further
reconstruction of quantum mechanics can be proceeded as in Hardy's
work~\cite{Hardy}. In particular, the reconstruction of
higher-dimensional systems from the two-dimensional ones and the
general transformations of the state after measurement are
explicitly given there. We only briefly comment on them here.

In order to derive the state space, measurements and transformations
for a higher-dimensional system, we can use quantum theory of a
two-dimensional system  in conjunction with axiom 1. The axiom
requires that upon any two linearly independent states one can
construct a two-dimensional subspace that is isomorphic to the state
space of a qubit (2-sphere). The state space of a higher dimensional
system can be characterized such that if the state is restricted to
any given two dimensional subspace, then it behaves like a qubit.
The fact that all other (higher-dimensional) systems can be built
out of two-dimensional ones suggests that the latter can be
considered as fundamental constituents of the world and gives a
justification for the usage of the term ``elementary system'' in the
formulations of the axioms.

When a measurement is performed and an outcome is obtain, our
knowledge about the state of the system changes and its
representation in form of the  probabilities must be updated to be
in agreement with the new knowledge acquired in the measurement.
This is the most natural update rule present in any probability
theory. Only if one views this change as a real physical process conceptual problems arise related to discontinuous and abrupt ``collapse of the wave function''. There is no basis for any such assumption. Associated with each outcome is the measurement vector $p$.
When the outcome is observed the state after the measurement is
updated to $p$ and the measurement will be a certain transformation
on the initial state. Update rules for more general measurements can
accordingly be given.

\section{What the present reconstruction tells us about quantum mechanics}

It is often said that reconstructions of quantum theory within an
operational approach are devoid of ontological commitments, and that
nothing can be generally said about the ontological content that
arises from the first principles or about the status of the notion
of realism. As a supporting argument one usually notes that within a
realistic world view one would anyway expect quantum theory at the
operational level to be deducible from some underlying theory of
``deeper reality''. After all, we have the Broglie-Bohm
theory~\cite{Bohm} which is a nonlocal realistic theory in full
agreement with the predictions of (non-relativistic) quantum theory.
Having said this, we cannot but emphasize that realism does stay
``orthogonal'' to the basic idea behind our reconstruction.

Be it local or nonlocal, realism asserts that outcomes correspond to
actualities objectively existing prior to and independent of
measurements. On the other hand, we have shown that the finiteness of information
carrying capacity of quantum systems is an
important ingredient in deriving quantum theory. This capacity is not enough to allow 
assignment of definite values to outcomes of all possible measurements. The elementary system has the information carrying capacity of one bit. 
This is signified
by the possibility to decompose any state of an elementary system
(qubit) in quantum mechanics in two orthogonal states. In a
realistic theory based on hidden variables and an ``epistemic
constraint'' on an observer's knowledge of the variables’ values one
can reproduce this feature at the level of the entire distribution
of the hidden variables~\cite{Spekkens2}. That this is possible is
not surprising if one bears in mind that hidden-variable theories
were at the first place introduced to {\em reproduce} quantum
mechanics and yet give a more complete description~\footnote{That
this cannot be done without allowing nonlocal influences from
space-like distant regions is a valid point for itself, which we do
not want to follow here further.}. But any realism of that kind at the same time
assumes an {\em infinite information capacity} at the level of
hidden variables. Even to reproduce measurements on a single qubit
requires infinitely many orthogonal hidden-variable
states~\cite{Hardy2,Montina,Dakic}.
It might be a matter of taste whether or not one is ready to work
with this ``ontological access baggage``~\cite{Hardy2} not doing any
explanatory work at the operational level. But it is certainly
conceptually distinctly different from the theory analyzed here, in
which the information capacity of the most elementary systems  --
those which are by definition not reducible further -- is
fundamentally limited.

To further clarify our position consider the Mach-Zehnder
interferometer in which both the path information and interference
observable are dichotomic, i.e. two-valued observables. It is meaningless to speak about ``the path the
particle took in the interferometer in the interference
experiment'' because this would already require to assign 2 bits of
information to the system, which would exceed its information
capacity of 1 bit~\cite{Young}. The information capacity
of the system is simply not enough to provide definite
outcomes to all possible measurements. Then, by necessity the
outcome in some experiments must contain an element of randomness
and there must be observables that are complementarity to each
other. Entanglement and consequently the violation of Bell's
inequality (and thus of local realism) arise from the possibility to
define an abstract elementary system carrying at most one bit such
that correlations (``00'' and ``11'' in a joint measurement of two
subsystems) are basis states.


\section{Conclusions}

Quantum theory is our most accurate description of nature and is
fundamental to our understanding of, for example, the stability of matter, the
periodic table of chemical elements, and the energy of the sun. It has
led to the development of great inventions like the electronic
transistor, the laser, or quantum cryptography. Given the enormous
success of quantum theory, can we consider it as our final and
ultimate theory? Quantum theory has caused much controversy in
interpreting what its philosophical and epistemological implications
are. At the heart of this controversy lies the fact that the theory
makes only probabilistic predictions. In recent years it was however
shown that some features of quantum theory that one might have
expected to be uniquely quantum, turned out to be highly generic for
generalized probabilistic theories. Is there any reason why the
universe should obey the laws of quantum theory, as opposed to any
other possible probabilistic theory?

In this work we have shown that classical probability theory and
quantum theory -- the only two probability theories for which we
have empirical evidences --- are special in a way that they fulfill
three reasonable axioms on the systems' information carrying
capacity, on the notion of locality and on the reversibility of
transformations. The two theories can be separated if one restricts
the transformations between the pure states to be
continuous~\cite{Hardy}. An interesting finding is that quantum
theory is the {\it only} non-classical probability theory that can
exhibit entanglement without conflicting one or more axioms.
Therefore -- to use Schr\"{o}dinger's words~\cite{Schroedinger} --
entanglement is not only ``{\it the} characteristic trait of quantum
mechanics, the one that enforces its entire departure from classical
lines of thought'', but also the one that enforces the departure
from a broad class of more general probabilistic theories.

\acknowledgements

We thank M. Aspelmeyer, J. Kofler, T. Paterek and A. Zeilinger for discussions.
We acknowledge support from the Austrian Science Foundation FWF within
Project No. P19570-N16, SFB and CoQuS No. W1210-N16, the European Commission
Project QAP (No. 015848) and the Foundational Question Institute (FQXi).

\section{Appendix}

In this appendix we give the proofs of the lemmas from the main
text.

\begin{lemma}The lower bound $||T||=1$ is saturated, if and only if
the state is a product state $T=\bx\by^{\mathrm{T}}$.
\end{lemma}

{\bf Proof.} If the state is a product state then
$||T||^2=||\bx||^2||\by||^2=1$. On the other hand, assume that the
state $\psi=(\bx,\by,T)$ satisfies $||T||=1$. Normalization
(\ref{norm}) gives $||\bx||=||\by||=1$. Let
$\phi_p=(-\bx,-\by,T_0=\bx\by^{\mathrm{T}})$ be a product state. We have
$P(\psi,\phi_p)\geq0$ and therefore
\begin{eqnarray}
1-||\bx||^2-||\by||^2+\mathrm{Tr}(T^{\mathrm{T}}T_0)=-1+\mathrm{Tr}(T^{\mathrm{T}}T_0)\geq0.
\end{eqnarray}
The last inequality $\mathrm{Tr}(T^{\mathrm{T}}T_0)\geq1$ can be seen as
$(T,T_0)\geq1$ where $(,)$ is the scalar product in Hilbert-Schmidt
space. Since the vectors $T,T_0$ are normalized, $||T||=||T_0||=1$,
the scalar product between them is always $(T,T_0)\leq1$. Therefore,
we have $(T,T_0)=1$ which is equivalent to $T=T_0=\bx\by^{\mathrm{T}}$.
\begin{flushright}
QED
 \end{flushright}

\begin{lemma} The only product states belonging to $S_{12}$ are $\psi_1$ and $\psi_2$.
\end{lemma}

{\bf Proof.} Let $\psi_p=(\bx,\by,\bx\by^{\mathrm{T}})\in S_{12}$. We
have
\begin{eqnarray}
1&=&P_{12}(\psi_p,\psi_1)+P_{12}(\psi_p,\psi_2)\\
&=&\frac{1}{4}(1+\bx\be_1+\by\be_1+(\bx\be_1)(\by\be_1)) \\
&+& \frac{1}{4}(1-\bx\be_1-\by\be_1+(\bx\be_1)(\by\be_1))\\
&=&\frac{1}{2}(1+(\bx\be_1)(\by\be_1))\\
&\Rightarrow&\bx\be_1=\by\be_1=1~\vee~\bx\be_1=\by\be_1=-1\\
&\Leftrightarrow&\bx=\by=\be_1~\vee~\bx=\by=-\be_1.
\end{eqnarray}
\begin{flushright}
QED
\end{flushright}

\begin{lemma}\label{sub flip}If the state $\psi\in S_{12}$, then $\psi'=(R,\openone)\psi\in S_{34}$ and $\psi''=(\openone,R)\psi\in
S_{34}$.
\end{lemma}
{\bf Proof.} If $\psi\in S_{12}$ we have
\begin{eqnarray}
1&=&P_{12}(\psi,\psi_1)+P_{12}(\psi,\psi_2)\\
&=&P_{12}((R,\openone)\psi,(R,\openone)\psi_1)+P_{12}((R,\openone)\psi,(R,\openone)\psi_2)\\
&=&P_{12}(\psi',\psi_3)+P_{12}(\psi',\psi_4).
\end{eqnarray}
Similarly, one can show that $(\openone,R)\psi\in
S_{34}$.
\begin{flushright}
QED
\end{flushright}

\begin{lemma}\label{partial transpose} Let $U$ be some operator with the following action in the Hilbert-Schmidt space; $U(\rho)=U\rho U^{\dagger}$,
and $PT_1$ and $PT_2$ are partial transpositions with respect to
subsystems 1 and 2, respectively. The following identity holds:
$\mathrm{PT}_1U\mathrm{PT}_1=\mathrm{PT}_2U^{\ast}\mathrm{PT}_2$,
where $U^{\ast}$ is the complex-conjugate operator.
\end{lemma}
{\bf Proof.} We can expand $U$ into some product basis in the
Hilbert-Schmidt space $U=\sum_{ij}u_{ij}A_i\otimes B_j$. We have
\begin{eqnarray}
\mathrm{PT}_1U\mathrm{PT}_1(\rho_1\otimes\rho_2)&=&\mathrm{PT}_1\{U\rho_1^{\mathrm{T}}\otimes\rho_2U^{\dagger}\}\\\nonumber
&=&\sum_{ijkl}u_{ij}u^{\ast}_{kl}(A_k^{\ast}\rho_1A_i^{\mathrm{T}})\otimes(B_j\rho_2B_l^{\dagger})\\\nonumber
&=&\mathrm{PT}_2\{\sum_{ijkl}u_{ij}u^{\ast}_{kl}(A_k^{\ast}\rho_1A_i^{\mathrm{T}})\otimes(B_l^{\ast}\rho_2^{\mathrm{T}}B_j^{\mathrm{T}})\}\\\nonumber
&=&\mathrm{PT}_2\{\sum_{ijkl}u_{kl}^{\ast}u_{ij}(A_k^{\ast}\otimes
B_l^{\ast})(\rho_1\otimes\rho_2^{\mathrm{T}})(A_i^{T}\otimes
B_j^{T})\}\\\nonumber
&=&\mathrm{PT}_2U^{\ast}\mathrm{PT}_2(\rho_1\otimes\rho_2),
\end{eqnarray}
for arbitrary operators $\rho_1$ and $\rho_2$.

\begin{flushright}
QED
\end{flushright}


\begin{thebibliography}{99}


%



\bibitem{KS} S. Kochen and E.P. Specker, {\em The Problem of Hidden Variables in Quantum Mechanics}, J. Math. Mech. {\bf 17}, 59 (1967).

\bibitem{Bell} J.S. Bell, {\em On the Einstein-Podolsky-Rosen paradox}, Physics {\bf 1}, 195-200 (1964); reprinted in J.S. Bell, ``Speakable and Unspeakable in
Quantum Mechanics'' (Cambridge Univ. Press, Cambridge, 1987).

\bibitem{Leggett} A.J. Leggett, {\em Nonlocal Hidden-Variable Theories and Quantum Mechanics: An Incompatibility Theorem}, Found. Phys. {\bf 33}, 1469 (2003).

\bibitem{Groeblacher} S. Gr\"{o}blacher, T. Paterek, R. Kaltenbaek, {\v C}. Brukner, M. {\. Z}ukowski, M. Aspelmeyer and A. Zeilinger, {\em An experimental test of non-local
realism}, Nature {\bf 446}, 871 (2007).

\bibitem{Birula} I. Biaynicki-Birula and J. Mycielski, {\em Nonlinear Wave Mechanics}, Ann. Phys. {\bf 100}, 62 (1976).

\bibitem{Shimony} A. Shimony, {\em Proposed neutron interferometer test of some nonlinear variants of wave mechanics}, Phys. Rev. A {\bf 20}, 394 (1979).

\bibitem{Shull} C.G. Shull, D. K. Atwood, J. Arthur, and M. A. Horne, {\em Search for a Nonlinear Variant of the Schrödinger Equation by Neutron Interferometry}, Phys. Rev. Lett. {\bf 44}, 765 (1980).

\bibitem{Gaehler} R. G\"{a}hler, A.G. Klein, A. Zeilinger, {\em Neutron Optical Tests of Nonlinear Wave Mechanics}, Phys. Rev. A {\bf 23}, 1611 (1981).

\bibitem{GWR} G. C. Ghirardi, A. Rimini and T. Weber, {\em Unified dynamics for microscopic and macroscopic systems}, Phys. Rev. D {\bf 34}, 470 (1986).

\bibitem{Karoli} F. K\'{a}rolyh\'{a}zy, Nuovo Cimento {\bf 42}, 390 (1966). F. K\'{a}rolyh\'{a}zy, {\em Gravitation and Quantum Mechanics of Macroscopic Bodies} (Thesis, in Hungarian), Magyar Fizikai Foly\'{o}irat {\bf 22}, 23 (1974).

\bibitem{Diosi} L. Diosi, {\em Models for universal reduction of macroscopic quantum fluctuations}, Phys. Rev. A {\bf 40}, 1165 (1989).

\bibitem{Penrose} R. Penrose, {\em On Gravity's role in Quantum State Reduction}, Gen. Relativ. Gravit. {\bf 28}, 581 (1996).

\bibitem{Pearle} P. Pearle, {\em Reduction of the state vector by a nonlinear Schrödinger equation}, Phys. Rev. D {\bf 13}, 857 (1976).

\bibitem{Barnum06} H. Barnum, J. Barrett, M. Leifer and A. Wilce, {\em Cloning and broadcasting in generic probabilistic models}, (2006)
(arXiv:quant-ph/061129).

\bibitem{Barnum07} H. Barnum, J. Barrett, M. Leifer and A.Wilce, {\em A general no-cloning theorem}, Phys. Rev. Lett. {\bf 99}, 240501 (2007)
(arXiv:0707.0620).

\bibitem{Barrett} J. Barrett, {\em Information processing in general probabilistic theories}, Phys. Rev. A. {\bf 75}, 032304 (2007)
(arXiv:quant-ph/0508211).

\bibitem{PRBox} S. Popescu and D. Rohrlich, {\em Quantum nonlocality as an axiom},
Found. Phys. {\bf 24}, 379 (1994).

\bibitem{Mackey} G. W. Mackey, {\em Quantum Mechanics and Hilbert Space}, American
Mathematical Monthly {\bf 64}, 45 (1957).

\bibitem{Hardy} L. Hardy, {\em Quantum Theory From Five Reasonable Axioms} (2001) (arXiv.org/quant-ph/0101012).

\bibitem{Dariano} G.M. D’Ariano, {\em Probabilistic theories: what is special about Quantum Mechanics?}, in ``Philosophy of Quantum Information
and Entanglement'', Eds. A. Bokulich and G. Jaeger (Cambridge University Press, Cambridge UK), (arXiv:0807.4383).

\bibitem{Goyal2} P. Goyal, K.H. Knuth and J. Skilling, {\em Origin of Complex Quantum Amplitudes and Feynman's Rules} (2009) (arXiv:0907.0909)

\bibitem{Rovelli} C. Rovelli, {\em Relational Quantum Mechanics}, Int. J. Theor. Phys. {\bf 35}, 1637 (1996).

\bibitem{Zeilinger} A. Zeilinger, {\em A Foundational Principle for Quantum Mechanics}, Found. Phys. {\bf 29}, 631 (1999).

\bibitem{BruknerZeilinger} {\v C}. Brukner and A. Zeilinger, {\em Information and Fundamental Elements of the Structure of Quantum Theory}, in ``Time, Quantum, Information'', Eds. L. Castell and O. Ischebeck (Springer, 2003) (arXiv:quant-ph/0212084). {\v C}. Brukner and A. Zeilinger, {\em Information Invariance and Quantum Probabilities}, Found. Phys. {\bf 39}, 677 (2009).

\bibitem{CBH} R. Clifton, J. Bub, and H. Halvorson, {\em Characterizing Quantum Theory
in Terms of Information-Theoretic Constraints}, Found. Phys. {\bf 33}(11), 1561 (2003).

\bibitem{Grinbaum1} A. Grinbaum, {\em Elements of information-theoretic derivation of the formalism of quantum theory}, Int. J. Quant. Inf. {\bf 1}(3), 289 (2003).

\bibitem{Wootters} W.K. Wootters, {\em Statistical distance and Hilbert space}, Phys. Rev. D {\bf 23}, 357 (1981).

\bibitem{Fivel} D.I. Fivel, {\em How interference effects in mixtures determine the rules of quantum mechanics}, Phys. Rev. A {\bf 59}, 2108 (1994).

\bibitem{Summhammer} J. Summhammer, {\em Maximum predictive power and the superposition principle}, Int. J. Theor. Phys. {\bf 33}, 171 (1994). J. Summhammer, {\em Quantum Theory as Efficient Representation of Probabilistic Information}, (2007) (arXiv:quant-ph/0701181).

\bibitem{Bohr} A. Bohr and O. Ulfbeck, {\em Primary manifestation of symmetry. Origin of quantal indeterminacy}, Rev. Mod. Phys. {\bf 67}, 1 (1995).

\bibitem{Caticha} A. Caticha, {\em Consistency, amplitudes and probabilities in quantum theory}, Phys. Rev. A {\bf 57}, 1572 (1998).

\bibitem{Fuchs} C.A. Fuchs, {\em Quantum mechanics as quantum information (and only a little more)}, in Ed. A. Khrenikov ``Quantum Theory: Reconstruction
of Foundations'' (V\"{a}xjo, V\"{a}xjo University Press, 2002) (quant-ph/ 0205039). C.A. Fuchs and R. Schack, {\em Quantum-Bayesian Coherence}, (2009) (arXiv:0906.2187).

\bibitem{Grangier} P. Grangier, {\em Contextual objectivity : a realistic interpretation of quantum mechanics}, Eur. J. Phys. {\bf 23}, 331 (2002) (arXiv:quant-ph/0012122); P. Grangier, {\em Contextual objectivity and the quantum formalism}, Proc. of the conference ``Foundations of Quantum
Information'' (Camerino, Italy, 2004) (arxiv.org/quant-ph/0407025).

\bibitem{Luo} S. Luo, {\em Maximum Shannon Entropy, Minimum Fisher Information, and an Elementary Game}, Found. Phys. {\bf 32}, 1757 (2002).

\bibitem{Spekkens} R. Spekkens, {\em Evidence for the epistemic view of quantum states: A toy theory}, Phys. Rev. A {\bf 75}, 032110 (2007).

\bibitem{Goyal1} P. Goyal, {\em Information-geometric reconstruction of quantum theory}, Phys. Rev. A {\bf 78}, 052120 (2008).

\bibitem{Vandam} W. van Dam, {\em Implausible Consequences of Superstrong Nonlocality}, (2005) (arXiv:quant-ph/0501159).

\bibitem{Brassard}  G. Brassard, H. Buhrman, N. Linden, A. Methot, A. Tapp and F. Unger, {\em A limit on nonlocality in any world in which communication complexity is not trivial}, Phys.
Rev. Lett. {\bf 96}, 250401 (2006).

\bibitem{Pawlowski} M. Pawlowski, T. Paterek, D. Kaszlikowski, V. Scarani, A. Winter, and M. Zukowski, {\em  A new physical principle: Information Causality} (2009) (arXiv:0905.2292).

\bibitem{Navascues}  M. Navascues and H. Wunderlich, {\em A glance beyond the quantum model} (arXiv:0907.0372).

\bibitem{Zyczkowski} K. Zyczkowski, {\em Quartic quantum theory: an extension of the standard quantum mechanics}, J. Phys. A {\bf 41}, 355302 (2008).

\bibitem{Paterek} T. Paterek, B. Dakic, {\v C}. Brukner, {\em Theories of systems with limited information content} (2008) (arXiv:0804.1423).

\bibitem{Peres1} A. Peres, {\em Proposed test for complex versus quaternion quantum theory}, Phys. Rev. Lett. {\bf 42}, 683 (1979).

\bibitem{Kaiser} H. Kaiser, E.A. George, and S.A. Werner, {\em Neutron interferometric search
for quaternions in quantum mechanics}, Phys. Rev A {\bf 29}, 2276 (1984).

\bibitem{Peres2} A. Peres, {\em Quaternionic quantum interferometry}, in ``Quantum Interferometry'', Eds. F. De Martini et al., (VCH Publ., 1996), 431-437 (arXiv:quant-ph/9605024).

\bibitem{Sorkin} R. D. Sorkin, {\em Quantum Mechanics as Quantum Measure Theory}, Mod. Phys. Lett. A {\bf 9}, 3119 (1994) (arXiv:gr-qc/9401003).

\bibitem{Weihs} U. Sinha, C. Couteau, Z. Medendorp, I. Söllner, R. Laflamme, R. Sorkin and G. Weihs, {\em Testing Born's Rule in Quantum Mechanics with a Triple Slit Experiment}, (2008) (arXiv:0811.2068). Submitted to the proceedings of Foundations of Probability and Physics-5, Vaxjo, Sweden, August 2008.

\bibitem{Grinbaum2} A. Grinbaum, {\em Reconstruction of Quantum Theory}, Brit. J. Phil. Sci. {\bf 8}, 387 (2007).

\bibitem{Gross} D. Gross, M. Mueller, R. Colbeck, and O.C.O. Dahlsten, {\em All reversible dynamics in maximally non-local theories are trivial} (2009) (arXiv:0910.1840). O.C.O. Dahlstein, privite communication.

\bibitem{Weizsaecker} C.F. von Weizs\"{a}cker, 1958, {\em Aufbau der Physik} (Carl Hanser, M\"{u}nchen,1958).

\bibitem{Wheeler} J.A. Wheeler, {\em Law without Law in Quantum Theory and Measurement}, Eds. J.A. Wheeler and W.H.
Zurek (Princeton University Press, Princeton, 1983) 182.

\bibitem{Barnum3} H. Barnum and A. Wilce, {\em Information processing in convex operational theories}, to be published in DCM/QPL (Developments in Computational Models / Quantum Programming Languages) (Oxford University, 2009) (arXiv:0908.2352).

\bibitem{Abramsky} S. Abramsky and B. Coecke, {\em A categorical semantics of quantum
protocols}, Proc. 19th IEEE Conference on Logic in Computer Science, 415–425 (IEEE Computer Science Press, 2004).

\bibitem{Loll} See, for example, J. Ambjorn, J. Jurkiewicz and R. Loll, {\em Reconstructing the Universe}, Phys. Rev. D {\bf 72} 064014 (2005).

\bibitem{Boerner} H. Boerner, {\em Representations of groups}, (North- Holland publishing
company, Amsterdam 1963).

\bibitem{Perron-Frobenius} R.A. Horn and C.R. Johnson, {\em Matrix Analysis}, (Cambridge University
Press, Chapter 8, 1990).

\bibitem{Rob} R. Spekkens, privite communication.

\bibitem{Bohm} D. Bohm, {\em A Suggested Interpretation of the Quantum Theory in Terms of ``Hidden Variables'' I}, Phys. Rev. {\bf 85}, 166 (1952).
D. Bohm, {\em A Suggested Interpretation of the Quantum Theory in Terms of ``Hidden Variables'' II}, Phys. Rev. {\bf 85}, 180  (1952).

\bibitem{Spekkens2} See Ref~\cite{Spekkens} for a local version of such hidden-variable theory in which quantum mechanical predictions are partially reproduced.

\bibitem{Hardy2} L. Hardy, {\em Quantum Ontological Excess Baggage}, Stud. Hist. Philos. Mod. Phys. {\bf 35}, 267 (2004).

\bibitem{Montina} A. Montina, {\em Exponential growth of the ontological space dimension with the physical size}, Phys. Rev. A {\bf 77}, 022104 (2008).

\bibitem{Dakic} B. Dakic, M. Suvakov, T. Paterek, and {\v C}. Brukner, {\em Efficient Hidden-Variable Simulation of Measurements in Quantum Experiments}, Phys. Rev. Lett. {\bf 101}, 190402 (2008).

\bibitem{Young} {\v C}. Brukner and A. Zeilinger, {\em Young's experiment and the finiteness of information}, Phil. Trans. R. Soc. Lond. A {\bf 360}, 1061 (2002).

\bibitem{Schroedinger} E. Schr\"{o}dinger, {\em Discussion of Probability Relations Between Separated Systems}, Proceedings of the Cambridge Philosophical Society 31 (1935) 555-563; 32 (1936): 446-451


\end{thebibliography}
\end{document}